# ANTASID: A Novel Temporal Adjustment to Shannon's Index of Difficulty for Quantifying the Perceived Difficulty of Uncontrolled Pointing Tasks

Mohammad Ridwan Kabir[a, c, 1], Mohammad Ishrak Abedin[b, c], Rizvi Ahmed[b, c], Hasan Mahmud[a, c],
and Md. Kamrul Hasan[a, c, 2]

[a] *Systems and Software Lab (SSL)*

[b] *Network and Data Analysis Group (NDAG)*

[c] *Department of Computer Science and Engineering*

*Islamic University of Technology (IUT), Gazipur, Bangladesh*

{ridwankabir, ishrakabedin, rizviahmed, hasan, hasank}@iut-dhaka.edu

Shannon's Index of Difficulty ($ID$), reputable for quantifying the perceived difficulty of pointing tasks as a logarithmic relationship between *movement-amplitude* ($A$) and *target-width* ($W$), is used for modelling the corresponding *observed movement-times* ($MT_O$) in such tasks in controlled experimental setup. However, real-life pointing tasks are both spatially and temporally uncontrolled, being influenced by factors such as – human aspects, subjective behavior, the context of interaction, the inherent speed-accuracy trade-off where, emphasizing accuracy compromises speed of interaction and vice versa, and so on. Effective target-width ($W_e$) is considered as spatial adjustment for compensating accuracy. However, no significant adjustment exists in the literature for compensating speed in different contexts of interaction in these tasks. As a result, without any temporal adjustment, the true difficulty of an uncontrolled pointing task may be inaccurately quantified using Shannon's $ID$. To verify this, we propose ANTASID (A Novel Temporal Adjustment to Shannon's ID) formulation with detailed performance analysis. We hypothesized a temporal adjustment factor ($t$) as a binary logarithm of $MT_O$, compensating for speed and minimizing the non-linearity between *movement-amplitude* and *target-width* due to contextual differences. Considering spatial and/or temporal adjustments to $ID$, we conducted regression analysis using our own and *Benchmark* datasets in both controlled and uncontrolled scenarios of pointing tasks with a generic mouse. ANTASID formulation showed significantly superior fitness values and throughput in all the scenarios while reducing the standard error. Furthermore, the quantification of $ID$ with ANTASID varied significantly compared to the classical formulations of Shannon's $ID$, validating the purpose of this study.

CCS CONCEPTS • **Human-centered computing~Human computer interaction (HCI)~HCI theory, concepts and models** • **Human-centered computing~Human computer interaction (HCI)~Interaction techniques~Pointing** • **Human-centered computing~Human computer interaction (HCI)~Interaction devices~Pointing devices** • **Human-centered computing~Human computer interaction (HCI)~Empirical studies in HCI**

## 1 INTRODUCTION

In the domain of Human-Computer Interaction (HCI), a pointing task is considered as the user interaction for selecting any element on a user interface with any pointing device such as mouse, stylus, trackpad, finger, or any other wearable devices. *Shannon's Index of Difficulty* ($ID$) [1] (Equation 1) is reputable for quantifying the perceived difficulty of such a task. It is expressed as a logarithmic relation between *movement-amplitude* ($A$) and *target-width* ($W$), where $A$ is defined as the distance between the starting location of the cursor and the target's center. Applying Fitts's law, the *movement-time* ($MT$) of any pointing task can be modeled as a linear function of $ID$ (Equation 2), where the constants $a$ and $b$ are empirically defined from a regression analysis of experimental data. The ideology of $ID$ is that the difficulty of a task increases as $A$ increases and/or $W$ decreases. However, it fails to address the speed-accuracy trade-off in pointing tasks [2]. The basic idea of this trade-off is that if users prefer to be efficient in terms of speed, the *observed movement-time* ($MT_O$) will be shorter and if the focus is shifted to accuracy, it will be longer. To account for the variability in accuracy, researchers have formulated a *Spatially Adjusted* ($SA$) variant of $ID$ (Equation 3) by replacing the *nominal target-width*, $W$ with the *effective target-width*, $W_e$ [1], [3], [4]. $W_e$ can be calculated either using the standard deviation method given endpoint coordinates are recorded (Equation 4) or using the discrete-error method given the error-rates of pointing are recorded [1].

$$ID = \log_2\left(\frac{A}{W} + 1\right) \qquad (1)$$

---

[1] *Corresponding Author*
[2] *Grant Recipient from* IUT RSG, *Grant No.:* REASP/IUT-RSG/2021/OL/07/012.

$$MT = a + b * ID \qquad (2)$$

$$ID_{SA} = \log_2\left(\frac{A}{W_e} + 1\right) \qquad (3)$$

$$W_e = 4.133 * SD_x \qquad (4)$$

The index of performance, also known as throughput ($TP$) (Equation 5), is defined as the average of the ratio of $ID$ and $MT_O$ over $n$ pointing tasks [1], [5], where $TP$ increases proportionally with $ID$.

$$TP = \frac{1}{n}\sum_{i=1}^{n} \frac{ID_i}{MT_{O_i}} \qquad (5)$$

In most studies related to Fitts's law [2], [5]–[11], *controlled* experiments are conducted in a lab setup with either subjective or parametric (*manipulating A or W*), or operational constraints (*extremely accurate, accurate, neutral, fast,* and *extremely fast*). In *uncontrolled* experiments, however, no such constraints are imposed. The objective of *controlled* experiments is to understand the effect of manipulating a variable on other variables of interest. This is often preferred while exploring new aspects of HCI. As stated in literature [12], although these studies may have high internal validity, they are at a risk of low external validity. In other words, findings of these studies may not hold for a different experimental setup. Moreover, factors that are difficult to manipulate in *controlled* experiments, makes this issue even more complex [2]. Therefore, it is imperative to conduct *uncontrolled* experiments to understand the extent of validity of any theories or formulations. It is logical to consider pointing tasks in real-life as part of *uncontrolled* experiments as they are both spatially and temporally unconstrained and biased due to human factors such as physical inability, distraction, fatigue, excitement, cognition time, etc., thereby, introducing *context of interaction*.

To comprehend the speed-accuracy trade-off in real-life pointing tasks, let us consider a user in two different contexts of submitting an online exam script: 1) *well ahead of deadline* and 2) *at the verge of deadline*, with the click of a "*Submit*" button. In the first scenario, the person will be in a relaxed state of mind, naturally, s/he will not emphasize speed of interaction over accuracy. However, in the second scenario, the person will feel tensed, emphasizing speed of interaction over accuracy to meet the submission deadline. Therefore, the time taken to complete similar pointing tasks on the same interface, may vary depending on the context.

Moreover, interaction with two similar graphical interfaces for two different applications might be different based on the context and use cases of the applications. For example, *Facebook* being a "*Social Media*" platform and *Google Classroom* being an "*Educational*" platform, have a *"Post"* and a *"Submit"* button, respectively, for doing similar tasks, i.e., uploading contents to the platforms. However, while posting on *Facebook*, users are generally in a relaxed state of mind. On the contrary, while using *Google Classroom* as an exam script submission system, drawing from the previous example of submitting a script *at the verge of deadline*, the users are generally in a tensed state of mind. Therefore, in the case of *Facebook*, while a user will focus more on clicking the *"Post"* button *accurately* taking adequate time, in the other case, a user will focus more on clicking the *"Submit"* button quickly, minimizing the interaction time while emphasizing *speed* of interaction due to contextual differences.

It is evident from these discussions that the context of any pointing task affects the inherent speed-accuracy tradeoff along with the corresponding task-completion time. In such scenarios of uncontrolled pointing tasks, concerning *context of interaction* and subjective behavior [2], Shannon's $ID$ may fail to quantify the perceived difficulty of these tasks with or without spatial adjustment, justifying the need for a *temporal adjustment factor* ($t$). Although prior studies on Fitts's law conducted *controlled* experiments [2], [5]–[11] and *uncontrolled* experiments [13]–[17] with various adjustments to Shannon's $ID$, no significant evidence was found regarding *temporal adjustments* to account for the context and speed-accuracy tradeoff of interaction.

In this work, we present a novel formulation of a *temporal adjustment factor* ($t$) as the *binary logarithm* of *observed movement-time* ($MT_O$) of pointing tasks to quantify the contextual information of the task in *bits*. We have augmented the unadjusted (Equation 1) and the spatially adjusted (Equation 3) formulation of Shannon's $ID$ with $t$ as a power factor of $W$, to form ANTASID (*A Novel Temporal Adjustment* to *Shannon's ID*) formulation for quantifying $ID$ of such tasks. We hypothesized that Shannon's $ID$ may not be able to accurately quantify the perceived difficulty of pointing tasks due to the subjective and contextual behavior, and speed-accuracy trade-off with or without spatial adjustment. Hence, augmenting it with $t$ may resolve these issues and ensure a reliable quantification of $ID$.

To comprehend the significance of the *temporal adjustment factor* ($t$) in the aforementioned cases, we have conducted an *uncontrolled* experiment through a pointing-task-based game (developed in-house) using an optical mouse as a pointing device. We have defined *uncontrolled* pointing tasks as those where – 1) no operational constraints are imposed and 2) no manipulation of $A$ occurs. We have constructed an *Internal Dataset* with the *uncontrolled* user-interaction data from this game. The authors in [13] proposed a new derivation of Fitts's Law and conducted both *controlled* and *uncontrolled* experiments on pointing tasks. Datasets of both of their experiments [14] are publicly available. In addition to the data from the *uncontrolled* experiments of the *Benchmark* [14] and the *Internal Datasets*, we have leveraged the data from the *controlled* experiments of the *Benchmark Dataset* to verify the external validity of ANTASID formulation in *controlled* scenarios of pointing tasks as well. To summarize, all the formulations of Shannon's $ID$ considered in this study, were analyzed on – 1) two experimental datasets (*Internal* (6163 pointing tasks) and *Benchmark* (39050 pointing tasks)), featuring *uncontrolled* experiments and 2) one experimental dataset (*Benchmark* (8345 pointing tasks)), featuring *controlled* experiments.

To validate the significance of $t$ in quantifying $ID$ of pointing tasks in both controlled and uncontrolled scenarios and to verify how accurately the classical formulation of Shannon's $ID$ can quantify the perceived difficulty of these tasks, we analyzed $MT$ from four different regression models, considering *spatial* and/or *temporal* adjustments to Shannon's $ID$, such as- *neither spatially nor temporally adjusted* ($ID_{NA}$), *only temporally adjusted* ($ID_{TA}$), *only spatially adjusted* ($ID_{SA}$), and *both temporally and spatially adjusted* ($ID_{TSA}$). We state our null hypothesis as, $H_0$: *"Temporally adjusted Shannon's $ID$ might not accurately quantify the perceived difficulty of pointing tasks in both controlled and uncontrolled scenarios."* To be able to reject $H_0$, we need to verify whether $ID_{TA}$ and $ID_{TSA}$ provide a better model fit and are statistically significant over $ID_{NA}$ and $ID_{SA}$ in both *controlled* and *uncontrolled* scenarios. From the statistical analyses of data using one-way ANOVA, one-way F-test followed by a post-hoc test, we have found that ANTASID formulation significantly enhances the performance of the regression models in both scenarios of pointing tasks, validating the significance of $t$ in addressing the speed-accuracy trade-off in such tasks, considering subjective behavior and context of interaction using Fitts's law. The level of significance ($\alpha$) of the statistical analysis was considered as $\alpha$=0.05.

To summarize our contributions: 1) We have formulated a *temporal adjustment factor* ($t$) for better quantification of $ID$ by considering the context of interaction in real-life uncontrolled pointing tasks. 2) We have generated a dataset, containing the parameters required for quantifying $ID$ of uncontrolled pointing tasks using Fitts's law with a generic mouse. 3) We have analyzed the statistical significance of ANTASID as well as the classical formulation of Shannon's $ID$ and verified that the classical ones might not accurately quantify $ID$ for uncontrolled pointing tasks based on the context of interaction.

In the next sections, we present the literature review followed by an explanation of the proposed ANTASID formulation in details. We then elaborate on the user study followed by the result analysis section and a discussion on the properties of ANTASID formulation. Finally, we summarize our observations and give a direction on future works.

## 2  LITERATURE REVIEW

Researchers have been trying to understand the impact of speed-accuracy trade-off in pointing tasks on Fitts's law for quite a while. The correct formulation of $ID$ given the nature and constraints of pointing tasks is still an active research area.

Having analyzed the impact of speed-accuracy trade-off in pointing tasks in [2], the authors have proposed a modified spatial adjustment factor, $W_m$ (Equation 6), where $\alpha$ is a power factor expressing a nonlinear relation between $A$ and $W_m$. $ID$ was formulated using this modified spatial adjustment factor $W_m$ (Equation 7). The value of $\alpha$ was empirically determined and cannot be generalized for other datasets. The authors imposed five operational constraints (*extremely accurate, accurate, neutral, fast,* and *extremely fast*) in their experiment. In our experiment, however, we did not impose any operational constraints of such sort to get an understanding of how Fitts's law performs in uncontrolled experiments using the classical as well as the proposed formulation of $ID$.

$$W_m = W \left(\frac{4.133 * SD_x}{W}\right)^{\alpha} \qquad (6)$$

$$ID = \log_2 \left(\frac{A}{W_m} + 1\right) \qquad (7)$$

A human motor behavioral model in distal pointing tasks [6] has been explored that formulates $ID$ as a function of angular amplitude ($\alpha$), angular target width ($\omega$) and an empirically defined constant ($k$) (Equation 8). The authors considered $k$ as a power of $\omega$ due to the nonlinear relationship between $\alpha$ and $\omega$. However, they did not provide any mathematical derivation of $k$. As a result, $R^2$ value of the regression model varied for different values of $k$. Using their proposed formulation of $ID$, they achieved an $R^2$ value of 0.961 for $k=3$ in a *controlled* study.

Researchers have also studied the effect of screen size variations on $ID$ in *controlled* experimental conditions [7]. Their findings revealed that the ratio of $A$ and $W$ in $ID$, fails to capture the true perceived difficulty in pointing tasks. They proposed modifications in the formulation of $ID$ for larger (Equation 9) and smaller (Equation 10) screen sizes. The terms $\alpha$ and $\beta$ were empirically determined. Motivated by their work, we have defined the temporal adjustment factor ($t$) as a power of $W$. However, rather than quantifying an experiment-specific value of $t$ we have defined it as a function of $MT_O$ of each pointing task for better quantification of $ID$ in any experimental setup.

$$ID = \left[\log_2\left(\frac{\alpha}{\omega^k} + 1\right)\right]^2 \quad (8)$$

$$ID = \log_2\left(\frac{A^\alpha}{W} + 1\right), \alpha > 1 \quad (9)$$

$$ID = \log_2\left(\frac{A}{W^\beta} + 1\right), \beta > 1 \quad (10)$$

A predictive error model was derived in [8] by manipulating parameters of Fitts's law such as, $W, A$, and $MT_O$. The authors reported that $W$ has a greater influence on error-rate than $A$. They also reported a logarithmic speed-accuracy trade-off described by Fitts's law.

In [9], the authors have analyzed the speed-accuracy phenomenon of Fitts's law in trajectory-based tasks with temporal constraints. They reported that in spatially constrained tasks, lateral deviation of the trajectory was affected by $W$ and subjective bias. On the other hand, in temporally constrained tasks, it was affected by $W$ and average steering speed.

Researchers in [10] reported that temporal constraint influences the speed-accuracy trade-off in aimed hand movements. Furthermore, the SH-model for pointing tasks was introduced based on the temporal distribution of successful hits and general principles of information theory [11]. The performance of this model was validated with the help of *AIC* (*Akaike's Information Criterion*) [18].

A new derivation of Fitts's law was also proposed based on velocity profile in pointing tasks [13]. The authors have compared their model's performance with widely accepted models. They conducted both *controlled* and *uncontrolled* experiments using *homogeneous* targets (the same value of $W$ for a group of targets) and *heterogeneous* targets (different value of $W$ for each target).

Considering maximum entropy, researchers have also explored an *Exponentially Modified Gaussian* (*EMG*) model to estimate a linear bound of linear regression in presence of outliers [15]. According to their work, data from such tasks have high variance and positive skewness, resulting in a very poor fit of the classical linear regression models. Considering their observations, we conducted experiments involving uncontrolled pointing tasks and analyzed the performance of various $ID$ formulations on the experimental data.

Considering uncontrolled aimed movements in *Graphical User Interfaces* (*GUI*), a real-life "*in the wild*" or in other words, an uncontrolled scenario of pointing tasks was analyzed by logging mouse cursor trajectories without imposing any constraints on user interaction [17]. The authors have introduced a spatial adjustment factor *Length Distance Index* (*LDI*) (Equation 11) in the formulation of $ID$ where, $L$ is the amplitude of movement and $D$ is the straight-line distance between the starting and the end points of the movement. Apart from the studies mentioned above, a formal information theoretic approach of resolving speed-accuracy tradeoff has also been explored [16].

$$LDI = \left(\frac{L}{D} - 1\right)^{\frac{1}{4}} \quad (11)$$

It is evident from the literature review that the relative weights of *movement-amplitude* ($A$) and *target-width* ($W$) have been adjusted by a power factor while quantifying $ID$, minimizing the inherent non-linearity between the two parameters. However, all these factors were empirically defined, and cannot be generalized for other experiments. In the next section, we discuss about the proposed ANTASID formulation, where we define the *temporal adjustment factor* ($t$) for analyzing the perceived difficulty of real-life pointing tasks applying Fitts's law.

## 3 ANTASID FORMULATION

ANTASID formulation introduces a *temporal adjustment factor* ($t$) to $ID$ as a power factor of $W$. The *Path Efficiency* ($PE$) of a pointing task is defined as the ratio *Straight Line Distance* ($SLD$) between the cursor position at the beginning of a task and the target's center to the *movement-amplitude* ($A$) [19], quantifying the corresponding *spatial efficiency*. Analogous to $PE$, our proposed adjustment factor, $t$, for a particular pointing task is based on the *Temporal Efficiency* ($TE$) of that task.

Taking the influence of external factors such as- context of interaction, biased human behavior, speed-accuracy trade-off, human factors, and so on [2], [5], [14], [6]–[13] into account, if there are $n$ pointing tasks in an experiment, we define the $TE$ of the $i^{th}$ pointing task ($TE_i$) as the ratio of the *average observed movement-time* ($\overline{MT_O}$) over the $n$ pointing tasks to the *observed movement-time* ($MT_O^i$) (Equation 12) of that task. From the data of our *uncontrolled* pointing task experiment, and that of the *controlled* and *uncontrolled* pointing experiments in the *Benchmark Dataset* [14], the value of $\overline{MT_O}$ was found to be 0.8282, 0.7628, and 0.8618 seconds for $n = 6163$, $n = 8345$, and $n = 39050$ pointing tasks, respectively. Based on this empirical data, we considered the value of $\overline{MT_O}$ as 1 second for mathematical convenience. However, in case of different experiments, for instance, analysis of Fitts's law in pointing tasks using wearable pointing device, if the value of $\overline{MT_O}$ deviates far from 1 second, the actual value of $\overline{MT_O}$ might produce better quantification of $ID$, subject to further investigation. Since $ID$ is expressed in *bits*, we defined $t_i$ of the $i^{th}$ pointing task as the binary logarithm of $TE_i$ (Equation 13), quantifying the *temporal* information of that task in *bits*. As $\overline{MT_O} = 1$, the expression $t_i$ (Equation 13) reduces to the negative binary logarithm of $MT_O^i$ (Equation 14) as $\log_2 1 = 0$.

$$TE_i = \frac{\overline{MT_O}}{MT_O^i} \ where, 1 \leq i \leq n \quad (12)$$

$$t_i = \log_2\left(\frac{\overline{MT_O}}{MT_O^i}\right), where \ \overline{MT_O} = 1 \ second \ and \ 1 \leq i \leq n \quad (13)$$

$$t_i = -\log_2(MT_O^i) \quad (14)$$

Shannon's $ID$ in its original form (Equation 1) is neither spatially nor temporally adjusted. From this point onward, we will refer to it as $ID_{NA}$ (Equation 15). The spatially adjusted $ID$ (Equation 3) will be referred to as $ID_{SA}$ (Equation 16). To address the nonlinear relationship between $A$ and $W$, most studies have adjusted $W$ with a power factor [2], [6]. Furthermore, as the movement time for any pointing task depends on two sub-movements [8], [9]: 1) Initial Ballistic Phase and 2) Optional Correction Phase, it is intuitive that for a target with smaller $W$, more time will be spent on the latter. Based on this intuition and evidence from the literature, our proposed ANTASID formulation augments $ID_{NA}$ and $ID_{SA}$ with $t$ as a power factor of $W$ to formulate $ID_{TA}$ (Equation 17) and $ID_{TSA}$ (Equation 18), respectively, where $ID_{TA}$ is only temporally adjusted and $ID_{TSA}$ is both temporally and spatially adjusted.

$$ID_{NA} = \log_2\left(\frac{A}{W} + 1\right) \quad (15)$$

$$ID_{SA} = \log_2\left(\frac{A}{W_e} + 1\right) \quad (16)$$

$$ID_{TA} = \log_2\left(\frac{A}{W^t} + 1\right) \quad (17)$$

$$ID_{TSA} = \log_2\left(\frac{A}{W_e^t} + 1\right) \quad (18)$$

Due to the speed-accuracy trade-off phenomenon accompanied by context of interaction and subjective behavior, by definition, the value of $t$ will be different for each of the $i \in n$ tasks and participants, resulting in a realistic value of $ID$. In the next section, we discuss about our experimental design and data analysis to investigate the validity of our formulation.

## 4  USER STUDY

To investigate the significance of ANTASID formulation in accurately quantifying $ID$, in both *controlled* and *uncontrolled* pointing tasks, we conducted an *uncontrolled* within-subject experiment, featuring a *balloon popping* game, "*Popper*" (implemented in-house using *Python*) with an *optical mouse* as a pointing device.

We have constructed an *Internal Dataset* by compiling the user-interaction data from this game followed by data analysis. We also analyzed the data from the *controlled* and the *uncontrolled* experiment of the *Benchmark Dataset* [14] for this purpose. After analyzing all the associated datasets, we have gained valuable insights of the performance of ANTASID formulation in both scenarios of pointing tasks.

### 4.1 Participants

25 right-handed volunteers [16 males (64%), 9 females (36%); Mean age: 23.20±2.14 years] participated in our study. All of them had adequate experience of operating a computer with a mouse. They were recruited through known acquaintances and provided a verbal consent prior to their participation.

### 4.2 Experimental Setup

The participants were asked to play the entire game 3 times on a laptop with a screen resolution of 1920x1080 pixels using a generic computer mouse as a pointing device. We allotted 15 minutes per participant during which they were briefed about the semantics of the game, had a few trial runs followed by actual experiment. Data for each play of the game were automatically uploaded to our server. The participants were notified about the automated data collection prior to their participation and were assured of no *invasion of privacy* from our part.

### 4.3 Experimental Design

Since in this work, we are focusing on *uncontrolled* pointing tasks, unlike the authors in [2], we did not impose any constraints such as – *extremely accurate, accurate, neutral, fast,* and *extremely fast* on their interaction with the game. The participants had to pop a total of 90 balloons of 4 different widths, $W$ (32 px, 64 px, 96 px, and 128 px) as targets, only one at a time, appearing at random locations on the screen. The game was designed to be run in full-screen mode, ensuring full utilization of the screen resolution. The game featured 2 types of levels, *homogeneous* level, having a group of targets of similar width, and *heterogeneous* level, having a group of targets of randomized widths. There was a total of 5 levels in our game, 4 *homogeneous* levels, each with 15 targets of same $W$ and only one *heterogeneous* level with 30 targets of randomized $W$. Therefore, a total of 60 sequential *homogeneous* targets and 30 sequential *heterogeneous* targets were presented to a participant. A quantitative summary of the game levels is provided in Table 1. A brief resting period was allocated after each level in the form of a reward screen to reduce fatigue. The game ended when all the targets had been popped by a participant and the data was uploaded to our server for further analysis.

Table 1: Qualitative and quantitative summary of the game "Popper".

| Level Type | Level Number | Target Width, $w_i \in W$ (pixels) | Total targets | Target Distribution (%) |
|---|---|---|---|---|
| Homogeneous | 1 | 128 | 15 | 66.67% |
|  | 2 | 96 | 15 |  |
|  | 3 | 64 | 15 |  |
|  | 4 | 32 | 15 |  |
| Heterogeneous | 5 | Random [32/64/96/128] | 30 | 33.33% |
|  |  | Total Targets | 90 | 100% |

In our experiment, we define the popping of a target as a trial. A trial began with the appearance of a target and ended when a participant clicked inside a target. Clicks outside the target boundary were recorded as miss-clicks. Trials with miss-clicks were not rejected as artifacts. The screen coordinate of the cursor *at the beginning* and *at the end* of a trial were considered as the *starting* and the *ending* coordinates, respectively. However, a participant could move the cursor around the screen freely, before a target appeared and after it was clicked. Therefore, there was no fixed starting coordinate for the mouse cursor in any trial, rather the cursor coordinates at the time of appearance of a new target was considered as the *starting* coordinate. Due to this arrangement, it is intuitive that the *movement-amplitude*, $A$ will vary, ensuring no spatial constraint is imposed. The *movement-amplitude* ($A$) was calculated as the sum of the *Euclidean Distances* ($EDs$) between two consecutive coordinates in the *cursor trajectory* and the *observed movement-time* ($MT_O$) was recorded as the duration of a trial. A descriptive summary of the parameters that were recorded per trial is summarized in Table 2. From 25 participants after 3 plays of the game "*Popper*", about 6750 trials [4500 *homogeneous* (66.67%), 2250 *heterogeneous* (33.33%)] were registered.

Table 2: Descriptive summary of parameters, recorded per trial, for the game "Popper".

| Parameter | Value(s) / Representation/ Unit | Interpretation |
|---|---|---|
| Level Number | $x$ | Level Number (1 – 5) |
| Target-Width (W) | 32, 64, 96, and 128 (pixels) | Target-width ($W$) in pixels. |
| Starting Coordinate | $(x, y)$ | Coordinate of the cursor at the time of target appearance. |
| Ending Coordinate | $(x, y)$ | Coordinate of the cursor at the time of clicking inside the target. |
| Target Center | $(x, y)$ | Coordinate of the center of a target. |
| Movement Time ($MT_O$) | Seconds | Time required to click on a target from the moment it appeared on the screen. |
| Number of miss-clicks | $x$ | Number of clicks outside the target boundary. |
| Coordinates of cursor trajectory | $[(x_1, y_1), (x_2, y_2), ..., (x_n, y_n)]$ | List of coordinates in the cursor trajectory from the *Starting Coordinate* to the *Ending Coordinate*. Used in the calculation of *movement-amplitude* ($A$). |

## 4.4 Data Preprocessing

Data analysis was carried out using *Python*. Prior to analysis, trials with erroneous parameter values (e.g., $MT_O = 0$ seconds) due to system issues were removed from the *Internal Dataset*. This cleanup is required to avoid erroneous calculation of $t$ (Equation 14). For example, for any trial $i$, if $MT_O^i = 0$ (due to system error), then $t_i = -\log_2 0$ is *undefined*. Furthermore, trials having $MT_O$ beyond 3 *Standard Deviations* ($SD$) of the *mean observed movement-time* ($\overline{MT_O}$) were removed from this dataset, which we term as the $L_1$ *cleanup*. A summary of data distribution after each cleanup is depicted in Table 3. A remarkable insight from this two-level cleanup on the *Internal Dataset* is the relatively constant ratio of *homogeneous* and *heterogeneous* trials at each level of cleanup.

Table 3: Distribution summary of Trials in the Internal Dataset at different levels of cleanup.

| Cleanup Type | Nature of Experiment | Number of Trials Before Cleanup | Accept/Reject Ratio of Trials after Cleanup | | | | Ratio of Trial Type after Cleanup | | | |
|---|---|---|---|---|---|---|---|---|---|---|
| | | | Accepted[a] | | Rejected | | Homogeneous | | Heterogeneous | |
| | | | N | % | N | % | N | % | N | % |
| Removal of Erroneous Data | Uncontrolled | 6750 | 6505 | 96.37 | 245 | 3.63 | 4319 | 66.40 | 2186 | 33.60 |
| $L_1$ | | 6505 | 6469 | 99.45 | 36 | 0.55 | 4291 | 66.33 | 2178 | 33.67 |

[a] Trials having $MT_O$ within 3 SD of $\overline{MT_O}$ were accepted.

As mentioned earlier, the *Benchmark Dataset* [14] contained data from both *controlled* and *uncontrolled* experiments, featuring *homogeneous* and *heterogeneous* targets. However, in the experiments with *homogeneous* targets, the authors manipulated both the *movement-amplitude* and the *target-width* which goes against our experimental design. Therefore, in our analysis, we considered data from both experiments featuring *heterogeneous* targets only. Although the authors had already removed trial data with $MT_O$ beyond 3 $SD$ of $\overline{MT_O}$ from this dataset [8350 *controlled* and 39050 *uncontrolled* trials post-cleanup], we further removed trials with erroneous parameter values (e.g., $MT_O = 0$ seconds) from this dataset, obtaining 8345 *controlled* and 39050 *uncontrolled* trials.

For the *Internal Dataset*, we calculated $W_e$ following the standard deviation method (Equation 4) [1]. However, one shortcoming of the *Benchmark Dataset* is that neither the coordinates of target selection and its center nor the percentage of errors were recorded. Therefore, $W_e$ cannot be calculated using the *standard deviation* method (Equation 4). However, the *discrete-error* method of determining $W_e$ [1] can be applied by approximating a reasonable error-rate ∈ % where, for error-rates less than ∈ %, $W_e < W$ and vice versa and for error-rates equal to ∈ %, $W_e = W$. Therefore, for the *Benchmark Dataset*, we approximated ∈=3.883% and calculated the average effective target-width ($\overline{W_e}$), such that $\overline{W_e} = \overline{W}$ (Equation 19), where $n$ is the number of trials, $W_i$ is the target-width for trial $i$, and $z$ is the Z-score at the corresponding error-rate. $ID_{SA}$ and $ID_{TSA}$ were calculated for this dataset using the values of $\overline{W_e}$ as shown in Table 4.

$$\overline{W_e} = \frac{1}{n}\sum_{i=1}^{n} \frac{2.066}{z} * W_i \qquad (19)$$

Table 4: Comparison of Average effective target-width ($\overline{W_e}$) at the approximated error-rate ($\in$) and the average target-width ($\overline{W}$) in the *Benchmark Dataset*.

| Nature of Experiment | $\overline{W}$ (pixels) | Approximation of $\overline{W_e}$ (pixels) ($\in$=3.883 %)[a] |
|---|---|---|
| Controlled | 30.1780 | 30.1782 |
| Uncontrolled | 30.2105 | 30.2108 |

[a] $\in$ is the approximated error-rate.

## 5 RESULT ANALYSIS

Using regression analysis, we predicted $MT$ using the different formulations of $ID$ mentioned earlier ($ID_{NA}, ID_{SA}, ID_{TA}$, and $ID_{TSA}$). We conducted a one-way ANOVA followed by a paired F-test for analyzing the corresponding statistical significance of the quantification of $ID$ using classical and ANTASID formulations. Furthermore, to avoid the risk of *Type-I* errors, we conducted the *Tukey's HSD* post-hoc test [20] on our formulations. We did not perform any comparative performance analysis of our models with that of the *Benchmark Dataset* [14] because of the differences in the formulation of $ID$. We have used Shannon's $ID$ and proposed adjustments to it while they have used the *Square-Root Variant* of Fitts's law.

From the regression analysis of the *Internal Dataset* (Figure 1), it is evident that ANTASID formulation provides a reasonable fit ($R^2$-value) of the model with ($R^2_{TSA}$= 0.8405) (Figure 1d) or without ($R^2_{TA}$= 0.8177) (Figure 1b) spatial adjustment. Similar results were observed for both the *controlled* ($R^2_{TSA}$=0.9095, $R^2_{TA}$=0.8521) (Figures 2d and 2b, respectively) and the *uncontrolled* ($R^2_{TSA}$=0.8953, $R^2_{TA}$=0.8308) (Figures 3d and 3b, respectively) experiments of the *Benchmark Dataset* at the approximated error-rate, $\in$=3.883%. Evidently, $ID_{TSA}$ consistently follows a normal distribution (Figures 4d, 4h, and 4l). From one-way ANOVA, it was observed that ANTASID formulations had significantly higher F-statistics at $p$<0.001 in all the datasets compared to $ID_{NA}$ and $ID_{SA}$. The parameters of regression analysis of the four models on different datasets along with corresponding results of ANOVA test are summarized in Table 5.

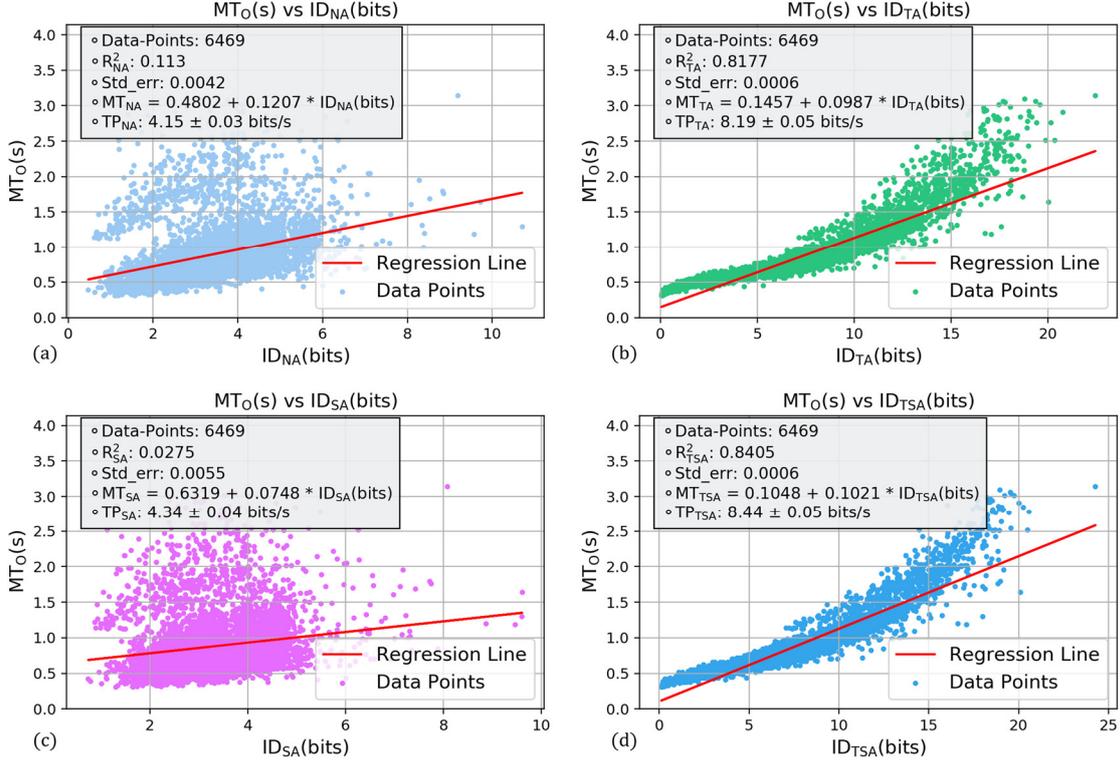

Figure 1: Regression Analysis of the Internal Dataset (Uncontrolled Experiment) using – (a) $ID_{NA}$, (b) $ID_{TA}$, (c) $ID_{SA}$, (d) $ID_{TSA}$.

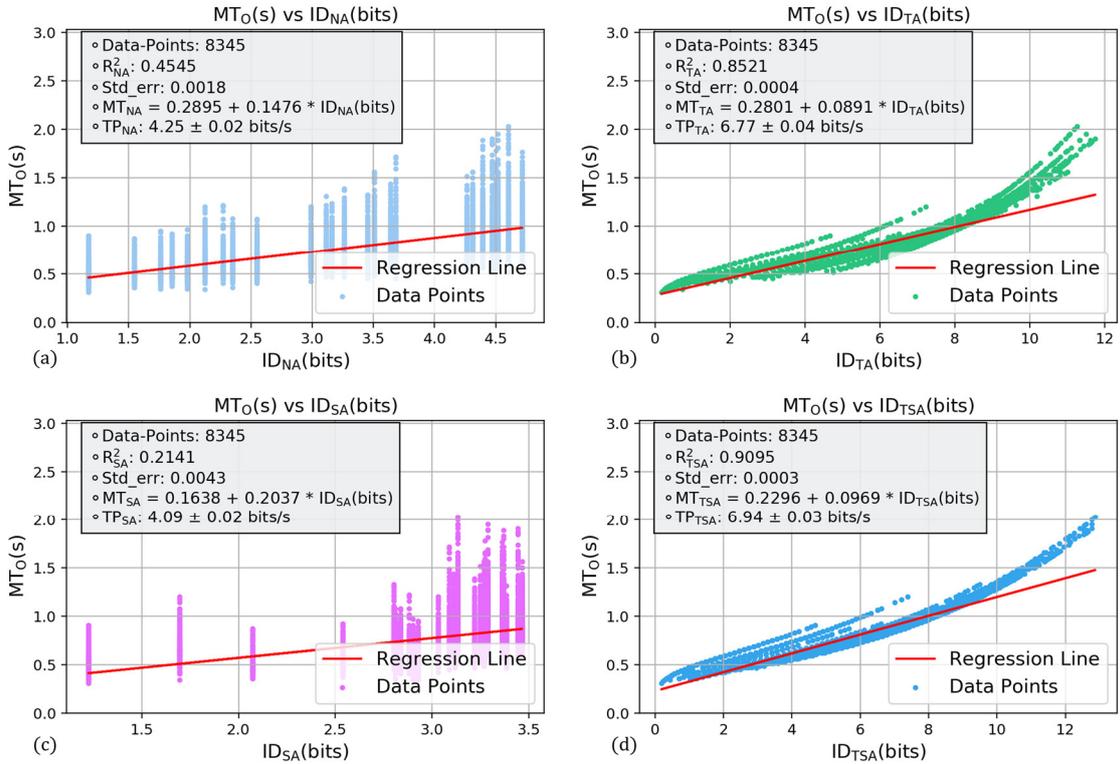

Figure 2: Regression Analysis of the Benchmark Dataset (Controlled Experiment) using – (a) $ID_{NA}$, (b) $ID_{TA}$, (c) $ID_{SA}$, (d) $ID_{TSA}$.

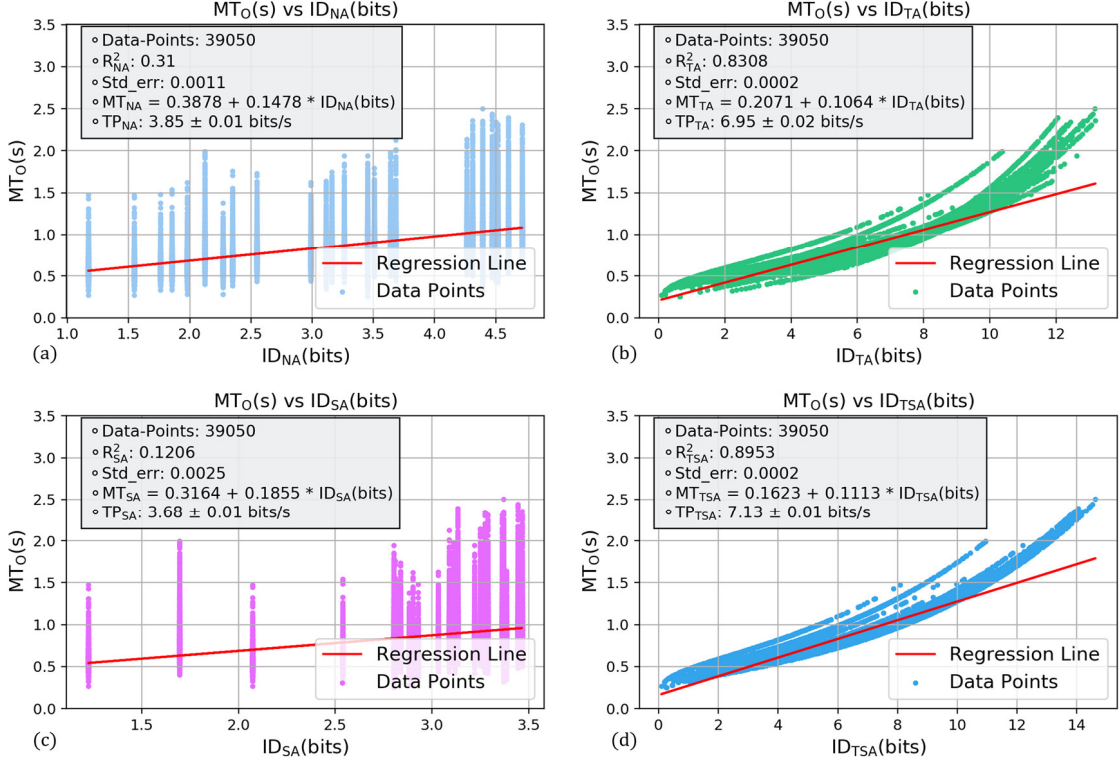

Figure 3: Regression Analysis of the Benchmark Dataset (Uncontrolled Experiment) using – (a) $ID_{NA}$, (b) $ID_{TA}$, (c) $ID_{SA}$, (d) $ID_{TSA}$.

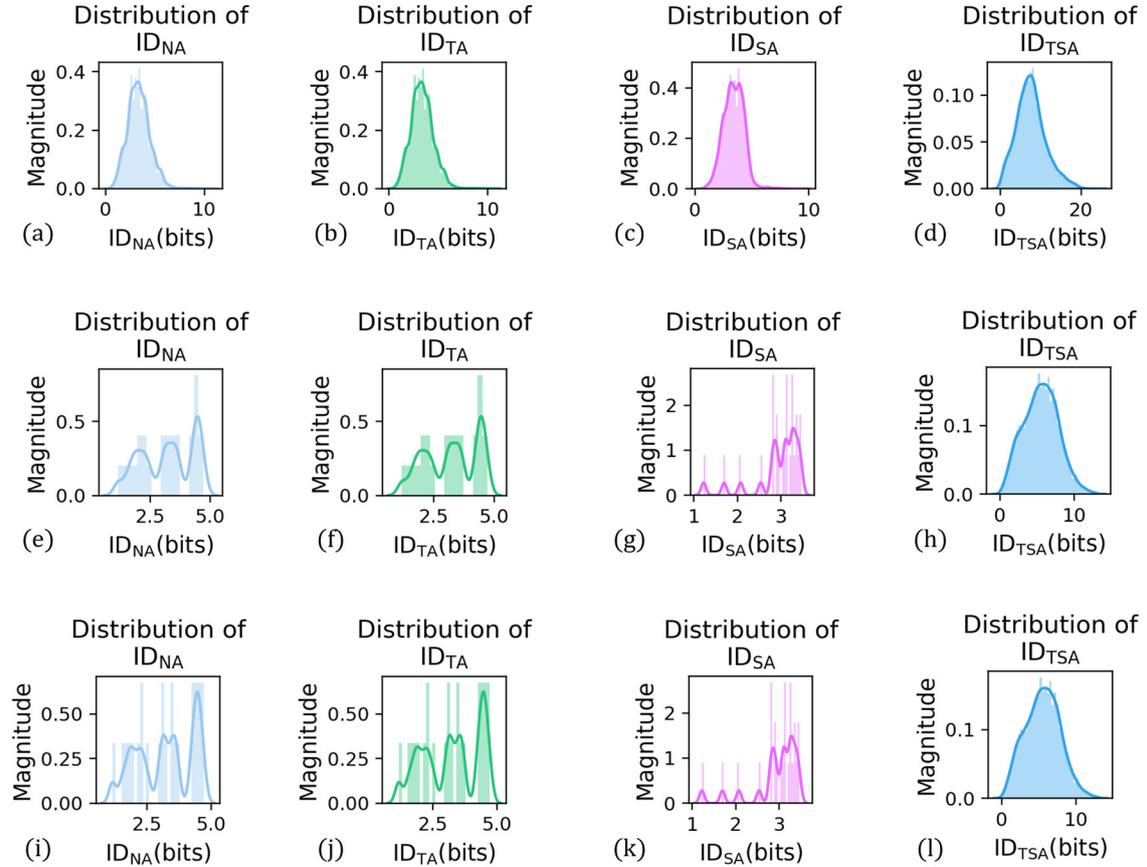

Figure 4: Distribution of different formulations of $ID$ – (a, b, c, d) Uncontrolled Experiment of the Internal Dataset. (e, f, g, h) Controlled Experiment of the Benchmark Dataset. (i, j, k, l) Uncontrolled Experiment of the Benchmark Dataset.

Given all the datasets, $ID_{TSA}$ has the best average model fit ($\overline{R^2}$=0.8818) along with the least *average standard error*, $\overline{SE}$=0.0004. Summary of the average performance of the models across all the datasets in Table 6 suggests that the *mean throughput*, $\overline{TP}$ almost doubles with ANTASID formulations compared to the classical ones. This is because, $TP$ increases continuously and proportionally with $ID_{TSA}$ and $ID_{TA}$, which can be visualized from the scatterplots of $TP$ vs $ID$ (Figure 5) of the *Internal Dataset*. Pairwise F-test as shown in Table 7 revealed the superiority of ANTASID formulation over both $ID_{NA}$ and $ID_{SA}$ at the desired level of significance, $\alpha$=0.05 with a *p*-value <0.001, in all the datasets considered in this study. The post-hoc test using Tukey's HSD [20] method also validated the same having an adjusted *p*-value of 0.001.

Table 5: Regression model parameters of Fitts's Law, $TP$ and ANOVA test results.

| Dataset (Experiment type) | Formulation Type | ID Formulation | $R^2$ value | Std. Error (SE) | $TP \pm$ 95%CI | ANOVA | | |
|---|---|---|---|---|---|---|---|---|
| | | | | | | DOF | F-stat | p-value[a] |
| Internal Dataset (Uncontrolled) | Classical | $ID_{NA}$ | 0.1130 | 0.0042 | 4.15±(0.03) | (1, 6467) | 824.12 | <0.001 |
| | | $ID_{SA}$ | 0.0275 | 0.0055 | 4.34±(0.04) | (1, 6467) | 182.78 | <0.001 |
| | ANTASID | $ID_{TA}$ | 0.8177 | 0.0006 | 8.19±(0.05) | (1, 6467) | 29010.66 | <0.001 |
| | | $ID_{TSA}$ | 0.8405 | 0.0006 | 8.44±(0.05) | (1, 6467) | 34087.95 | <0.001 |
| Benchmark Dataset (Uncontrolled) (∈=3.883%)[b] | Classical | $ID_{NA}$ | 0.3100 | 0.0011 | 3.85±(0.01) | (1, 39048) | 17541.32 | <0.001 |
| | | $ID_{SA}$ | 0.1206 | 0.0025 | 3.68±(0.01) | (1, 39048) | 5356.10 | <0.001 |
| | ANTASID | $ID_{TA}$ | 0.8308 | 0.0002 | 6.95±(0.02) | (1, 39048) | 191728.50 | <0.001 |
| | | $ID_{TSA}$ | 0.8953 | 0.0002 | 7.13±(0.01) | (1, 39048) | 334025.21 | <0.001 |
| Benchmark Dataset (Controlled) (∈=3.883%)[b] | Classical | $ID_{NA}$ | 0.4545 | 0.0018 | 4.25±(0.02) | (1, 8343) | 6950.55 | <0.001 |
| | | $ID_{SA}$ | 0.2141 | 0.0043 | 4.09±(0.02) | (1, 8343) | 2273.19 | <0.001 |
| | ANTASID | $ID_{TA}$ | 0.8521 | 0.0004 | 6.77±(0.04) | (1, 8343) | 48051.61 | <0.001 |
| | | $ID_{TSA}$ | 0.9095 | 0.0003 | 6.94±(0.03) | (1, 8343) | 83876.01 | <0.001 |

[a] *p*-values were computed at a significance level of, $\alpha$=0.05.
[b] ∈ is the approximated error-rate.

Table 6: Average performance of the regression models across all the datasets.

| Formulation Type | ID Formulation | $\overline{R^2}$ (±95% CI) | $\overline{SE}$ (±95% CI[a]) | $\overline{TP}$ (±95% CI[a]) (bits/s) |
|---|---|---|---|---|
| Classical | $ID_{NA}$ | 0.2925 (± 0.1227) | 0.0024 (± 0.0012) | 4.08 (± 0.15) |
| | $ID_{SA}$ | 0.1207 (± 0.0668) | 0.0041 (± 0.0011) | 3.03 (± 0.24) |
| ANTASID | $ID_{TA}$ | 0.8335 (± 0.0124) | 0.0004 (± 0.0001) | 7.30 (± 0.55) |
| | $ID_{TSA}$ | 0.8818 (± 0.0261) | 0.0004 (± 0.0001) | 7.50 (± 0.48) |

[a] CI: Confidence Interval

Table 7: Pairwise F-test results.

| ID Formulation ($\sigma_A^2 > \sigma_B^2$)[a] | | Internal Dataset (Uncontrolled Experiment) | | Benchmark Dataset (Uncontrolled Experiment) (∈=3.883 %)[b] | | Benchmark Dataset (Controlled Experiment) (∈=3.883 %)[b] | |
|---|---|---|---|---|---|---|---|
| A | B | F-stat | p-value[c] | F-stat | p-value[c] | F-stat | p-value[c] |
| $ID_{TSA}$ | $ID_{NA}$ | 7.4409 | <0.001 | 2.8881 | <0.001 | 2.0025 | <0.001 |
| $ID_{TSA}$ | $ID_{SA}$ | 30.6092 | <0.001 | 7.4203 | <0.001 | 4.2527 | <0.001 |
| $ID_{TA}$ | $ID_{NA}$ | 7.2310 | <0.001 | 2.6807 | <0.001 | 1.8744 | <0.001 |
| $ID_{TA}$ | $ID_{SA}$ | 29.7458 | <0.001 | 6.8873 | <0.001 | 3.9807 | <0.001 |

[a] Variance ($\sigma_i^2$) of $ID$, where $i \in \{A, B\}$.
[b] ∈ is the approximated error-rate.
[c] *p*-values were computed at a level of significance, $\alpha$=0.05.

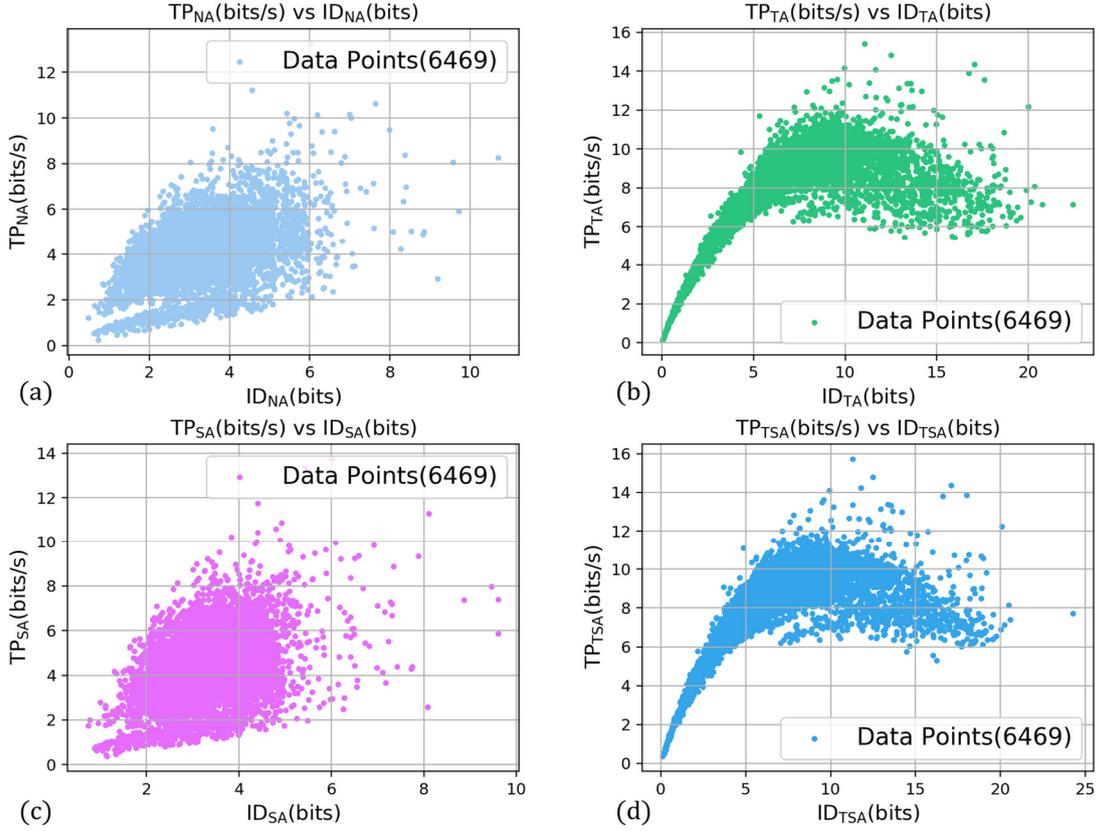

Figure 5: Analysis of the Throughput, *TP* corresponding to (a, c) the classical and (b, d) ANTASID formulations of Shannon's Index of Difficulty, *ID* for the Uncontrolled Experiment of the Internal Dataset.

These analyses imply that we can reject the null hypothesis, $H_0$ and accept the alternative hypothesis $H_1$:"*Temporally adjusted Shannon's ID may better quantify the perceived difficulty of pointing tasks in both controlled and uncontrolled scenarios.*" The *spatial adjustment* on top of the *temporal* one makes the formulation even more robust and provides a normally distributed *ID* along with enhanced *TP*.

## 6 DISCUSSIONS

Extensive studies have been carried out on Fitts's law over the years for understanding human performance in pointing tasks. These studies have proposed several variants of *ID* with the aim to develop an enhanced human interaction model. However, the perfect formulation of *ID* is still an active research area. In this work, we have proposed ANTASID formulation utilizing temporal efficiency of pointing tasks, reflecting the variation of perceived difficulty of pointing tasks in different contexts of interaction, and analyzed its effect alone or combined with *spatial adjustment* in the quantification of *ID* using an optical mouse. We have analyzed two separate datasets – 1) *Internal* dataset generated from our *uncontrolled* experiment of the game "*Popper*" and 2) *Benchmark* dataset featuring *controlled* and *uncontrolled* scenarios of pointing tasks. We have also statistically analyzed the performance of our formulation compared to the classical ones on these datasets at a significance level of $\alpha$=0.05

With respect to the experimental datasets considered in this study, the perceived difficulty of pointing tasks, quantified using ANTASID formulation ($ID_{TA}$, $ID_{TSA}$), has an enhanced inter-quartile range compared to the classical formulations of Shannon's *ID* ($ID_{NA}$, $ID_{SA}$) (Figure 6). The mean *predicted movement-times* using ANTASID ($\overline{MT_{TA}}, \overline{MT_{TSA}}$) and the classical formulations of Shannon's *ID* ($\overline{MT_{NA}}, \overline{MT_{SA}}$) are almost constant for both the *Internal* (Figure 7a) and the *Benchmark* (Figure 7b and 7c) datasets. Both $MT_{TA}$ and $MT_{TSA}$ exhibit interquartile ranges closer to the *observed movement-time* ($MT_O$), compared to both $MT_{NA}$ and $MT_{SA}$ (Figure 7). From this analysis, we can infer that the *temporal adjustment factor* ($t$) is able to capture the speed-accuracy trade-off phenomena of pointing tasks by adjusting the relative weights of $W$ and $A$ through ANTASID formulations ($ID_{TA}, ID_{TSA}$), exploiting the temporal efficiency of the user to reflect context-based deviation of perceived difficulty of the tasks. We have found this inference to be consistent across all the datasets.

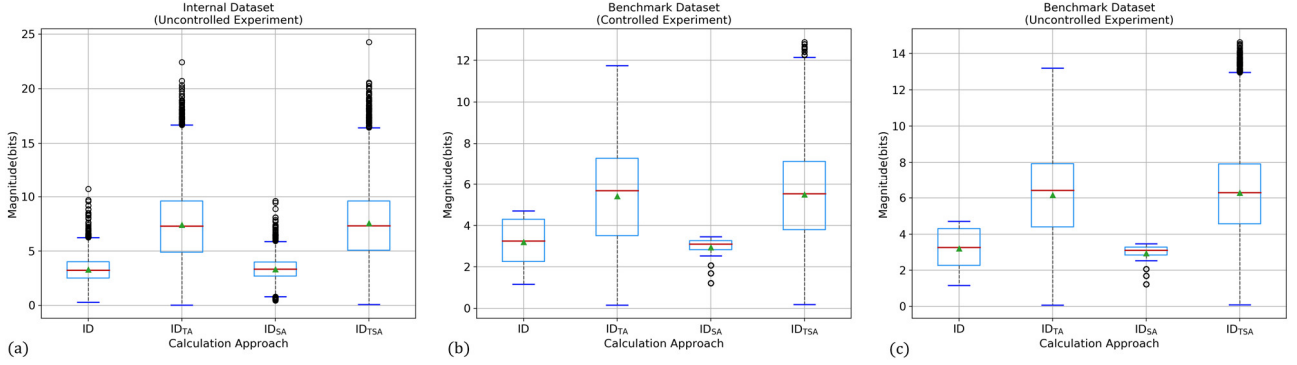

Figure 6: Boxplot analysis of the perceived difficulty of pointing tasks ($ID_{NA}, ID_{TA}, ID_{SA}, ID_{TSA}$) using ANTASID and the classical formulations of Shannon's Index of Difficulty considering – (a) the Uncontrolled Experiment of the Internal Dataset, (b) the Controlled and (c) the Uncontrolled Experiment of the Benchmark Dataset.

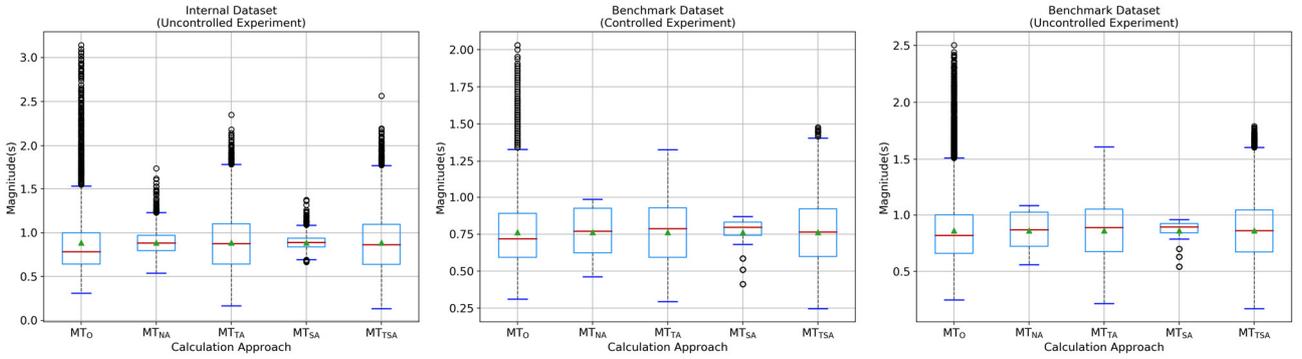

Figure 7: Boxplot analysis of the predicted movement-times ($MT_{NA}, MT_{TA}, MT_{SA}, MT_{TSA}$) using ANTASID and the classical formulations of Shannon's Index of Difficulty compared to the observed movement-time ($MT_O$) considering – (a) the Uncontrolled Experiment of the Internal Dataset, (b) the Controlled and (c) the Uncontrolled Experiment of the Benchmark Dataset.

A major observation from our analyses is that even in both *controlled* and *uncontrolled* scenarios of pointing tasks, ANTASID formulation significantly improved the fitness value of the regression model as well as $TP$ compared to its classical counterparts as seen from Table 6. This proves the robustness of ANTASID formulation in handling the context of interaction and the speed-accuracy trade-off in real-life pointing tasks. Evidently, ANTASID formulation is versatile and significantly overcomes the risk of low external validity.

We plan to extend the performance evaluation of ANTASID formulation to wearable pointing devices in both *controlled* and *uncontrolled* experimental procedures as part of our future work. This will help us identify the major factors that directly or indirectly influence the interaction of people with special needs with a computer using such devices. Furthermore, UIs of applications specific to certain contexts may be designed to meet specific index of difficulty ($ID$). Based on the desired $ID$, general context of interaction, and temporal efficiency of a specific pointing task, a possible practical significance of our study is the determination of a lower bound on the width ($W$) of targets (*windows, icons, menus, and navigation bars*) in the design of application UIs. This may not only help the UI designers in designing aesthetic UIs, but also enhance the throughput and efficiency of user interaction. Our proposed ANTASID formulation may be further used in testing the same UI in different circumstances. Taking context of interaction into account, the need for introducing a *temporal adjustment factor* ($t$) while quantifying the perceived difficulty of uncontrolled pointing task using Shannon's $ID$ cannot be over-emphasized, facilitating better comprehension of the efficiency of an interface with respect to its intended use case. For instance, in a click based real-time competitive game, though the UI remains same, considering different scenarios, a player might want to quickly perform certain interactions. For instance, in the beginning of real-time strategy games, generally the contention remains lesser than the middle or end portion of the game, where the players focus more towards quick decisions, actions and clicks. The proposed model can also be utilized in such scenarios to evaluate the perceived difficulty of real-time interaction under different context. We plan to carry out such research in the future, exploring the practical implications of the proposed formulation.

## ACKNOWLEDGMENTS

The authors express their heartfelt gratitude to the participants for their valuable time and effort for making this study possible. The authors do not declare any conflict of interest that may alter the outcomes of the study in any manner and approve this version of the manuscript for publication. by IUT RSG, Grant No.: REASP/IUT-RSG/2021/OL/07/012.

## REFERENCES

[1]  F. Law and I. S. Mackenzie, "Input / Output."

[2]  S. Zhai, J. Kong, X. R.-I. journal of human-computer studies, and undefined 2004, "Speed–accuracy tradeoff in Fitts' law tasks—on the equivalence of actual and nominal pointing precision," *Elsevier*, no. 6, pp. 823–856, 2004, Accessed: Dec. 28, 2021. [Online]. Available: https://www.sciencedirect.com/science/article/pii/S1071581904001028.

[3]  P. Fitts, J. P.-J. of experimental psychology, and undefined 1964, "Information capacity of discrete motor responses.," *psycnet.apa.org*, vol. 67, no. 2, 1964, Accessed: Dec. 28, 2021. [Online]. Available: https://psycnet.apa.org/record/1964-07019-001.

[4]  A. Welford, "Fundamentals of skill.," 1968, Accessed: Dec. 28, 2021. [Online]. Available: https://psycnet.apa.org/record/1968-35018-000.

[5]  S. Z.-I. J. of H.-C. Studies and undefined 2004, "Characterizing computer input with Fitts' law parameters—the information and non-information aspects of pointing," *Elsevier*, no. 6, pp. 791–809, 2004, doi: 10.1016/j.ijhcs.2004.09.006.

[6]  D. Bowman, R. Kopper, D. A. Bowman, M. G. Silva, and R. P. Mcmahan, "A human motor behavior model for distal pointing tasks," *Elsevier*, doi: 10.1016/j.ijhcs.2010.05.001.

[7]  H. Okada, T. A.-I. J. of C. and Information, and undefined 2014, "Evaluation of Fitts' Law Index of Difficulty Formulation for Screen Size Variations," *zenodo.net*, Accessed: Dec. 28, 2021. [Online]. Available: https://www.zenodo.net/record/1091304/files/9997694.pdf.

[8]  J. O. Wobbrock, E. Cutrell, S. Harada, and I. S. MacKenzie, "An error model for pointing based on Fitts' law," *Conf. Hum. Factors Comput. Syst. - Proc.*, pp. 1613–1622, 2008, doi: 10.1145/1357054.1357306.

[9]  X. Zhou, X. Cao, X. R.-I. C. on H.-C. Interaction, and undefined 2009, "Speed-accuracy tradeoff in trajectory-based tasks with temporal constraint," *Springer*, vol. 5726, pp. 906–919, 2009, Accessed: Dec. 28, 2021. [Online]. Available: https://link.springer.com/chapter/10.1007/978-3-642-03655-2_99.

[10]  H. Zelaznik, S. Mone, … G. M.-J. of E., and undefined 1988, "Role of temporal and spatial precision in determining the nature of the speed-accuracy trade-off in aimed-hand movements.," *psycnet.apa.org*, Accessed: Dec. 28, 2021. [Online]. Available: https://psycnet.apa.org/record/1988-25341-001.

[11]  X. Ren, J. Kong, X. J.-I. D. Courier, and undefined 2005, "SH-model: a model based on both system and human effects for pointing task evaluation," *jstage.jst.go.jp*, vol. 1, 2005, Accessed: Dec. 28, 2021. [Online]. Available: https://www.jstage.jst.go.jp/article/ipsjdc/1/0/1_0_193/_article/-char/ja/.

[12]  D. Gergle, D. T.-W. of K. in HCI, and undefined 2014, "Experimental research in HCI," *Springer*, pp. 191–227, Jan. 2014, doi: 10.1007/978-1-4939-0378-8_9.

[13]  K. Goldberg *et al.*, "A new derivation and dataset for Fitts' law of human motion," *eecs.berkeley.edu*, 2013, Accessed: Dec. 28, 2021. [Online]. Available: https://www2.eecs.berkeley.edu/Pubs/TechRpts/2013/EECS-2013-171.pdf.

[14]  K. Goldberg, S. Faridani, R. Alterovitz C A, and S. Francisco, "Two large open-access datasets for Fitts' law of human motion and a succinct derivation of the square-root variant," *ieeexplore.ieee.org*, Accessed: Dec. 28, 2021. [Online]. Available: https://ieeexplore.ieee.org/abstract/document/6994240/.

[15]  J. Gori, O. R.-2019 27th E. S. Processing, and undefined 2019, "Regression to a linear lower bound with outliers: An exponentially modified Gaussian noise model," *ieeexplore.ieee.org*, Accessed: Dec. 28, 2021. [Online]. Available: https://ieeexplore.ieee.org/abstract/document/8902946/.

[16]  J. Gori, O. Rioul, Y. G.-A. T. on Computer-Human, and undefined 2018, "Speed-accuracy tradeoff: A formal information-theoretic transmission scheme (fitts)," *dl.acm.org*, vol. 25, no. 5, Sep. 2018, doi: 10.1145/3231595.

[17]  O. Chapuis, R. Blanch, and M. Beaudouin-Lafon, "Fitts' law in the wild: A field study of aimed movements," 2007, Accessed: Dec. 28, 2021. [Online]. Available: https://hal.archives-ouvertes.fr/hal-00612026/.

[18]  D. Anderson, K. B.-J. of A. Statistics, and undefined 1998, "Comparison of Akaike information criterion and consistent Akaike information criterion for model selection and statistical inference from capture-recapture studies," *Taylor Fr.*, vol. 25, no. 2, pp. 263–282, 1998, doi: 10.1080/02664769823250.

[19]  A. HeydariGorji, S. Safavi, … C. L.-2017 I. S., and undefined 2017, "Head-mouse: A simple cursor controller based on optical measurement of head tilt," *ieeexplore.ieee.org*, Accessed: Dec. 28, 2021. [Online]. Available: https://ieeexplore.ieee.org/abstract/document/8257058/.

[20]  H. Abdi, L. W.-E. of research design, and undefined 2010, "Tukey's honestly significant difference (HSD) test," *researchgate.net*, 2014, Accessed: Dec. 28, 2021. [Online]. Available: https://www.researchgate.net/profile/Lynne-Williams-2/publication/237426041_Tukey%27s_Honestly_Signiflcant_Difierence_HSD_Test/links/00463528e752ddb7f3000000/Tukeys-Honestly-Signiflcant-Difierence-HSD-Test.pdf.